# Thermal Conductivity in the Bose-Einstein Condensed State of Triplons in the Bond-Alternating Spin-Chain System $Pb_2V_3O_9$

Mitsuhide Sato[1], Takayuki Kawamata[2], Naoki Sugawara[1], Naoto Kaneko[1], Masanori Uesaka[1], Kazutaka Kudo[3], Norio Kobayashi[3], Yoji Koike[1]

[1] Department of Applied Physics, Tohoku University, 6-6-05 Aoba, Aramaki, Aoba-ku, Sendai 980-8579, Japan
[2] Nishina Center for Accelerator-Based Science, RIKEN, 2-1 Hirosawa, Wako 351-0198, Japan
[3] Institute for Materials Research, Tohoku University, 2-1-1 Katahira, Aoba-ku, Sendai 980-8577, Japan

E-mail: mitsuhide@teion.apph.tohoku.ac.jp

**Abstract.** In order to clarify the origin of the enhancement of the thermal conductivity in the Bose-Einstein Condensed (BEC) state of field-induced triplons, we have measured the thermal conductivity along the [101] direction parallel to spin-chains, $\kappa_{\parallel[101]}$, and perpendicular to spin-chains, $\kappa_{\perp[101]}$, of the $S = 1/2$ bond-alternating spin-chain system $Pb_2V_3O_9$ in magnetic fields up to 14 T. With increasing field at 3 K, it has been found that both $\kappa_{\parallel[101]}$ and $\kappa_{\perp[101]}$ are suppressed in the gapped normal state in low fields. In the BEC state of field-induced triplons in high fields, on the other hand, $\kappa_{\parallel[101]}$ is enhanced with increasing field, while $\kappa_{\perp[101]}$ is suppressed. That is, the thermal conductivity along the direction, where the magnetic interaction is strong, is markedly enhanced in the BEC state. Accordingly, our results suggest that the enhancement of $\kappa_{\parallel[101]}$ in the BEC state is caused by the enhancement of the thermal conductivity due to triplons on the basis of the two-fluid model, as in the case of the superfluid state of liquid $^4$He.

## 1. Introduction

Recently, interesting magnetic states have been found to be induced by the application of magnetic field, pressure and so on in low-dimensional quantum spin systems. The thermal conductivity in such quantum spin systems has attracted considerable interest, because it often changes drastically due to the appearance of a field-induced new state. The thermal conductivity in low-dimensional quantum-spin insulators is given by the contribution of the thermal conductivity due to phonons, $\kappa_{phonon}$, and due to magnetic excitations, $\kappa_{spin}$. Therefore, the magnetic-field effect of the thermal conductivity appears through not only $\kappa_{spin}$ but also $\kappa_{phonon}$ via phonon-magnetic excitation scattering. For example, it has been reported that the thermal conductivity of $TlCuCl_3$ is drastically enhanced in the Bose-Einstein Condensed (BEC) state of field-induced magnetic excitations, namely, triplons [1]. However, it has not yet been clarified whether the enhancement in the BEC state is due to $\kappa_{phonon}$ or $\kappa_{spin}$.

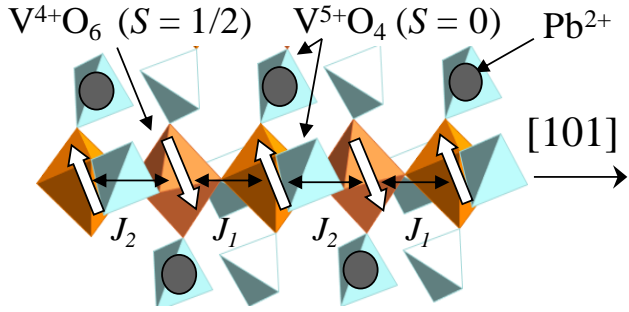

**Figure 1.** Crystal structure of $Pb_2V_3O_9$.

The one-dimensional (1D) quantum spin system $Pb_2V_3O_9$ is also known to give rise to a BEC state of field-induced triplons in high magnetic fields at low temperatures [2]. As shown in Fig. 1, this compound contains three kinds of vanadium ions in the unit cell. Two of them are non magnetic $V^{5+}$ ions located in $VO_4$ tetraheda, while the rest is $V^{4+}$ ions with the spin quantum number $S = 1/2$ in $VO_6$ octahedra. $VO_6$ octahedra are connected with each other by sharing an oxygen at the corner along the [101] direction, forming a $S = 1/2$ spin-chain. However, the magnitude of the magnetic interaction between adjacent spins is alternating, because the distance between adjacent spins along the [101] direction is alternating. Magnetic properties of $Pb_2V_3O_9$ are well understood in terms of the $S = 1/2$ 1D antiferromagnetic bond-alternating Heisenberg chain model. The magnetic interaction between adjacent spins is estimated as $J_1 = -29.2$ K and $J_2 = -19.3$ K from the magnetic susceptibility measurements [3]. The magnetic excitations are gapped at low temperatures and the spin gap, $\Delta$, is estimated as 7 K from the magnetization curve [2].

The anisotropy of the magnetic interaction of the 1D bond-alternating spin system $Pb_2V_3O_9$ is stronger than that of the three-dimensional spin-dimer system $TlCuCl_3$. Considering that the value of $\kappa_{spin}$ is related to the magnitude of the magnetic interaction and that $\kappa_{phonon}$ is comparatively isotropic, the origin of the enhancement of the thermal conductivity in the BEC state of field-induced triplons might be understood in terms of the anisotropy of the thermal conductivity in the BEC state of $Pb_2V_3O_9$. Therefore, we have measured the thermal conductivity of $Pb_2V_3O_9$ in magnetic fields parallel and perpendicular to spin-chains in order to investigate the enhancement of the thermal conductivity in the BEC state of field-induced triplons.

## 2. Experimental

Large-size single-crystals of $Pb_2V_3O_9$ were grown by the floating-zone method [4]. Thermal conductivity measurements were carried out by the conventional steady-state method. One side of a rectangular single-crystal was anchored on the copper heat sink with indium solder. A chip resistance (Alpha Electronics Corp., MP10K00) was attached as a heater to the opposite side of the single crystal with GE7031 varnish. The temperature difference across the crystal was measured with two Cernox thermometers (LakeShore Cryotronics, Inc., CX-1050-SD). Magnetic fields up to 14 T were applied parallel to the heat current, using a superconducting magnet.

## 3. Results and Discussion

Figure 2 shows the temperature dependence of the thermal conductivity along the [101] direction parallel to spin-chains, $\kappa_{\parallel[101]}$, and perpendicular to the [101] direction, $\kappa_{\perp[101]}$, which is perpendicular to the [101] direction and $b^*$-axis, in magnetic fields parallel to the respective heat current. It is found that both $\kappa_{\parallel[101]}$ and $\kappa_{\perp[101]}$ in the gapped normal state in zero field show a sharp peak around 9 K and the peak is suppressed with increasing field. The peak is due to $\kappa_{\mathrm{phonon}}$, because the thermal conductivity due to triplons, $\kappa_{\mathrm{triplon}}$, is very small on account of little population of triplons in the gapped normal state. The suppression of the peak by the application of magnetic field is attributable to the enhancement of the phonon-triplon scattering because of the reduction of the spin gap with increasing field. Actually, a similar behavior is observed in other spin-gap systems such as TlCuCl$_3$ [1,5] and SrCu$_2$(BO$_3$)$_2$ [5,6]. Such markedly enhancement as observed in TlCuCl$_3$ [1,5] is observed in the temperature dependence of neither $\kappa_{\parallel[101]}$ nor $\kappa_{\perp[101]}$ in the BEC state of Pb$_2$V$_3$O$_9$. However, it is found that $\kappa_{\parallel[101]}$ is slightly enhanced at low temperatures in high fields.

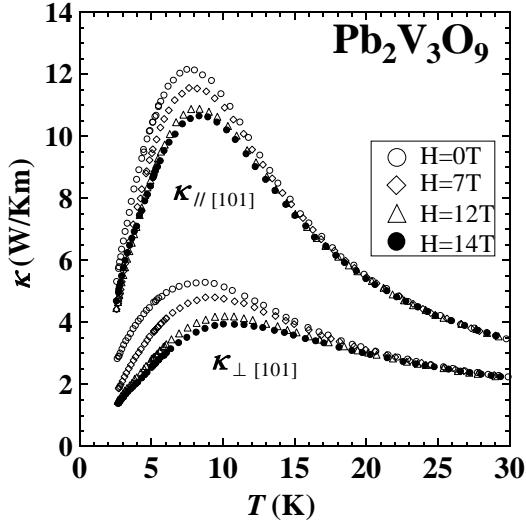
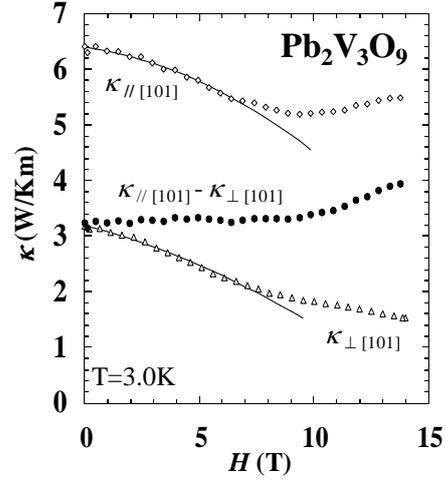

**Figure 2.** Temperature dependence of the thermal conductivity along the [101] direction parallel to spin-chains, $\kappa_{\parallel[101]}$, and perpendicular to the [101] direction, $\kappa_{\mathrm{perp}[101]}$, in magnetic fields parallel to the respective heat current.

**Figure 3.** Magnetic-field dependence of the thermal conductivity along the [101] direction parallel to spin-chains, $\kappa_{\parallel[101]}$, and perpendicular to [101] direction, $\kappa_{\mathrm{perp}[101]}$, in magnetic fields parallel to the respective heat current at 3.0 K. Closed circles are $\kappa_{\parallel[101]} - \kappa_{\mathrm{perp}[101]}$. Solid lines are the best-fit results between 0 T and 7 T using a parabolic field-dependence.

Figure 3 shows the magnetic-field dependence of $\kappa_{\parallel[101]}$ and $\kappa_{\perp[101]}$ at 3.0 K. It is found that both $\kappa_{\parallel[101]}$ and $\kappa_{\perp[101]}$ are suppressed with increasing field up to ~ 7 T and the suppression of the thermal conductivity is relaxed from ~ 7 T up to ~ 9 T. In high fields above ~ 9 T, $\kappa_{\parallel[101]}$ is enhanced, while $\kappa_{\perp[101]}$ is suppressed monotonically. The $\kappa_{\parallel[101]}$ is given by the summation of $\kappa_{\mathrm{phonon}}$ and $\kappa_{\mathrm{triplon}}$, while $\kappa_{\perp[101]}$ is given by only $\kappa_{\mathrm{phonon}}$. This is because $\kappa_{\mathrm{triplon}}$ is expected to be observed only along the direction parallel to the spin-chains where the magnetic interaction is strong. The magnetic field 7 T is in good correspondence with the transition field to the BEC state at 3 K [2]. Therefore, the relaxation of the suppression of $\kappa_{\perp[101]}$ above ~ 7 T is attributed to the decrease of the phonon-triplon scattering and understood based on the two-fluid model in the BEC state of triplons.

That is, it is interpreted as being due to the decrease of the population of uncondensed normal triplons bringing about the phonon-triplon scattering owing to the increase of the population of BEC triplons. As for the enhancement of $\kappa_{\parallel[101]}$ above ~ 9 T, it is probably interpreted as being due to the contribution of $\kappa_{triplon}$, because the phonon-triplon scattering affecting $\kappa_{phonon}$ is not simply expected to give rise to such a large anisotropic field-dependence of the thermal conductivity above ~ 9 T as observed in Fig. 3.

Here, we estimate the magnetic-field dependence of $\kappa_{triplon}$ at 3.0 K. Assuming that the anisotropy of $\kappa_{phonon}$ is very small, the contribution of $\kappa_{phonon}$ to $\kappa_{\parallel[101]}$ is given by $\kappa_{\perp[101]}$. Thus, $\kappa_{triplon}$ is simply calculated to be $\kappa_{\parallel[101]} - \kappa_{\perp[101]}$, as shown in Fig. 3. Taking into account the real anisotropy of $\kappa_{phonon}$, absolute values of $\kappa_{triplon}$ obtained thus are not exact, but anomalous dependence of $\kappa_{triplon}$ on magnetic field is able to be detected. It is found that $\kappa_{triplon}$ is constant up to ~ 9 T and increases with increasing field above ~ 9 T, which is larger than the transition field to the BEC state at 3 K, namely, ~ 7 T. The $\kappa_{triplon}$ is expected to decrease in the BEC state because of the decrease of the population of uncondensed normal triplons carrying heat. On the contrary, $\kappa_{triplon}$ is expected to increase in the BEC state because of the suppression of the scattering between uncondensed normal triplons. Therefore, it is likely that no field-dependence of $\kappa_{triplon}$ between ~ 7 T and ~ 9 T is due to balance between the two mechanisms. That is, the suppression of $\kappa_{triplon}$ is comparable to the enhancement of $\kappa_{triplon}$ in magnetic fields between ~ 7 T and ~ 9 T. As to the enhancement of $\kappa_{triplon}$ above ~ 9 T, there are two interpretation. First, it is interpreted as being due to the suppression of the scattering between uncondensed normal triplons in the BEC state, indicating that the effect of the suppression of the scattering is larger than that of the decrease of the population of uncondensed normal triplons. Second, it is interpreted as being due to the slightly different field dependence of $\kappa_{phonon}$, which might arise from an anisotropic phonon spectrum and from anisotropic scattering. The former interpretation is similar to the case of the thermal conductivity of the BEC superfluid state of liquid $^4$He [7]. Accordingly, we conclude that the enhancement of $\kappa_{\parallel[101]}$ in the BEC state of field-induced triplons in $Pb_2V_3O_9$ is attributed to the contribution of $\kappa_{triplon}$ owing to the similar mechanism as in the BEC superfluid state of liquid $^4$He. Additionally, it is strongly suggested that the enhancement of the thermal conductivity observed in the BEC state of field-induced triplons in $TlCuCl_3$ is also due to the contribution of $\kappa_{triplon}$.

## 4. Summary

In order to clarify the origin of the enhancement of the thermal conductivity in the BEC state of field-induced triplons, both $\kappa_{\parallel[101]}$ parallel to spin-chains and $\kappa_{\perp[101]}$ perpendicular to spin-chains of the $S = 1/2$ 1D bond-alternating spin system $Pb_2V_3O_9$ have been measured in magnetic fields up to 14 T parallel to the respective heat current. It has been found that $\kappa_{\parallel[101]}$ increases and $\kappa_{\perp[101]}$ decreases with increasing field above ~ 9 T in the BEC state. Considering the large anisotropy of the magnetic interaction, it is very likely that $\kappa_{\parallel[101]}$ is enhanced in the BEC state due to the contribution of $\kappa_{triplon}$ as in the case of the BEC superfluid state of liquid $^4$He. This result strongly suggests that the enhancement of the thermal conductivity in the BEC state of field-induced triplons is caused by the contribution of uncondensed normal triplons to the thermal conductivity.

**Acknowledgments**

The thermal conductivity measurements were performed at the High Field Laboratory for Superconducting Materials, Institute for Materials Research, Tohoku University. This work was partly supported by a Grant-in-Aid for Scientific Research from the Ministry of Education, Culture, Sports, Science and Technology, Japan.